\documentclass[letterpaper,twocolumn,10pt]{article}
\usepackage{usenix}

\usepackage{tikz}
\usepackage{amsmath}

\usepackage{booktabs}
\usepackage{tabularx}
\usepackage{threeparttable}
\usepackage{makecell}
\usepackage{adjustbox}
\usepackage{multirow}

\usepackage{array}

\usepackage{placeins}

\usepackage[most]{tcolorbox}

\begin{document}
%-------------------------------------------------------------------------------

%don't want date printed
\date{}

% make title bold and 14 pt font (LaTeX default is non-bold, 16 pt)
\title{SoK: Assessing Post-Quantum Cryptography Adoption in Software Development: Standards, Lifecycle Coverage, Deployment Maturity, and Evidence of Readiness}

%for single author (just remove % characters)
\author{
R.D.N. Shakya,$^{1,2}$ \qquad
Craig Costello,$^{3}$ \qquad
Nalin A.G. Arachchilage$^{2}$ \\
$^{1}$Faculty of Information Technology, University of Moratuwa, Sri Lanka \\
$^{2}$School of Computing Technologies, RMIT University, Australia \\
$^{3}$Queensland University of Technology, Australia \\
\texttt{shakyardn.26@uom.lk,craig.costello@qut.edu.au,nalin.arachchilage@rmit.edu.au}
}

\maketitle
%-------------------------------------------------------------------------------
\begin{abstract}
%-------------------------------------------------------------------------------

Standardised post-quantum cryptography (PQC) is critical for protecting software systems against future quantum threats, but standardised algorithms alone do not provide a pathway for their secure integration into software. This paper presents a Systematisation of Knowledge (SoK) of PQC integration in software development, synthesising existing approaches across implementation, migration, and developer education. We analyse these approaches against NIST PQC standards and PQC migration guidance, software development lifecycle coverage, cybersecurity practices, deployment maturity, and testing and evaluation evidence. Our analysis shows that existing work establishes important technical and organisational foundations but provides disconnected support across the software development lifecycle, with limited coordination between preparation, implementation, migration, deployment, maintenance, and developer practices. Evaluation evidence also remains concentrated in controlled or early-stage settings, providing stronger evidence of technical feasibility than operational readiness. We identify cross-cutting gaps and research directions for developing lifecycle-oriented, empirically validated approaches that connect technical, developer, and organisational requirements for secure PQC integration.

\end{abstract}

%-------------------------------------------------------------------------------
\section{Introduction}
\label{introduction}

Quantum computing poses a significant threat to widely deployed public-key cryptographic algorithms, including RSA and Elliptic Curve Cryptography, which secure modern software systems \cite{bernstein_post-quantum_2017, GILL20261}. Although cryptographically relevant quantum computers are not yet available, attackers may collect encrypted data today for future decryption, creating a "harvest now, decrypt later" threat \cite{bernardone2026quantum, giron_migrating_2023, Ana_huelsing2021wireguard}. Consequently, post-quantum cryptography (PQC) has become a strategic priority, shifting attention from developing quantum-resistant algorithms towards their practical integration and deployment in software systems \cite{bernstein_post-quantum_2017}.

In response, standards bodies and organisations worldwide have advanced PQC standardisation and deployment. The National Institute of Standards and Technology (NIST), for example, has standardised PQC algorithms through FIPS 203, FIPS 204, and FIPS 205 \cite{nist_pqc,nist_fips_203_2024,nist_fips_204_2024,nist_fips_205_2024}. However, standardisation alone does not resolve the engineering challenges of integrating PQC into software systems \cite{NISTTheoryReview}. Practical integration involves decisions concerning PQC algorithm selection and configuration, API and library integration, implementation, testing, validation, deployment, and maintenance \cite{howe_sok_2021, Ana_hekkala_implementing_2023, toruan2026security}. Flaws in these activities can introduce vulnerabilities or undermine the intended security guarantees \cite{howe_sok_2021, Ana_hekkala_implementing_2023, toruan2026security}. 

Thus, secure PQC integration requires complementary support, including frameworks, tools and libraries, guidance, migration strategies, and developer education \cite{Ana_hekkala_implementing_2023, ahmed_survey_2025, migration_Comparison, naether_migrating_2024, Ana_borrelli_towards_2025}. Collectively, these approaches form an emerging PQC integration landscape covering activities from implementing and integrating PQC mechanisms to migrating existing systems from classical cryptography. 

Despite this progress, the landscape remains fragmented \cite{anonymous_pqc_sok_2026}. Existing approaches \cite{Ana_huelsing2021wireguard, Ana_MigFrameOrg, Ana_MigFrDepen} often address individual activities, such as algorithm implementation, migration planning, or testing, rather than the end-to-end software development and deployment process. Addressing these activities in isolation can leave security-critical gaps; for example, effective migration planning does not ensure secure implementation, while technically correct implementation does not ensure that all vulnerable classical cryptographic dependencies have been identified and migrated.

It therefore remains unclear whether existing research provides sufficient support for the practical integration of standardised PQC into software systems. In particular, the extent to which existing approaches collectively cover software development lifecycle (SDLC) activities and align with PQC standards and migration recommendations has not been systematically examined. This is important because an approach may address a specific integration activity while relying on non-standardised algorithms, outdated implementations, or incomplete migration practices. Previous studies \cite{NISTTheoryReview, cryptography10010011, cryptography9020032, BenchAlgo} have primarily examined PQC algorithms or performance benchmarking and provide limited insight into how existing approaches collectively support PQC integration, including their alignment with standards and migration recommendations, SDLC and cybersecurity practice coverage, and deployment and evaluation evidence.

To address this gap, we present a Systematisation of Knowledge (SoK) of approaches supporting PQC integration in software. We systematically characterise existing approaches according to the aspects of PQC integration they support across software systems. Accordingly, we aim to systematically characterise the current PQC integration landscape in software development by addressing the following research questions (RQs):

\begin{itemize}
\item \textbf{RQ1:} What approaches have been proposed or developed to support PQC integration in software systems?
\item \textbf{RQ2:} To what extent do existing approaches align with available PQC standards and established migration recommendations?
\item \textbf{RQ3:} Which SDLC activities and cybersecurity activities are addressed by existing PQC approaches?
\item \textbf{RQ4:} What is the deployment maturity of existing approaches?
\item \textbf{RQ5:} What empirical evidence exists for the testing and evaluation of existing approaches? 
\end{itemize}

This SoK makes four primary contributions. First, we provide a systematic synthesis and classification of existing approaches supporting PQC implementation, migration, and developer education in software development, establishing a consolidated view of the current PQC integration landscape. Second, we present an evidence-driven assessment of these approaches by analysing their alignment with available PQC standards, migration recommendations, SDLC activities, and cybersecurity activities. Third, we evaluate the deployment maturity and empirical validation of existing approaches, revealing the gap between conceptual proposals and deployment-ready solutions. Finally, we identify key research challenges and future directions for advancing secure, scalable, and sustainable PQC integration in software development, thereby supporting the transition from standardised cryptographic algorithms to practical software deployment.

\section{Related work and Background}
\label{RWBack}

\subsection{Related work}
\label{RW}

A recent study examined socio-technical barriers affecting PQC implementation across human, technological, and organisational factors \cite{anonymous_pqc_sok_2026}. The present study extends this work by examining the broader PQC integration landscape across implementation, migration, and developer education approaches in software development. It further examines their alignment with PQC standards and migration guidance, coverage of software development lifecycle (SDLC) activities, deployment maturity, and evaluation evidence. Existing research has addressed specific aspects of this landscape. Ahmed et al.~\cite{ahmed_survey_2025} analyse PQC support in cryptographic libraries, while Naether et al.~\cite{naether_migrating_2024} examine software migration challenges. Bizri et al.~\cite{migration_Comparison} analyse institutional PQC migration frameworks. Other studies focus on algorithm foundations and implementations \cite{NISTTheoryReview}, PKI applications and benchmarking \cite{cryptography10010011}, and PQC performance and energy characteristics \cite{cryptography9020032, BenchAlgo}. Collectively, these studies provide insights into individual aspects of PQC integration but offer limited understanding of how these efforts can be connected to support secure PQC integration. This SoK addresses this gap by synthesising evidence on standards and migration-guidance alignment, SDLC coverage, cybersecurity activities, deployment maturity, and empirical evaluation.

\subsection{Background}
\label{Back}

To characterise the PQC integration landscape, this study adopts complementary reference \footnote{Only finalised and widely recognised standards, guidance documents, and recommendations were considered.} points discussed below. These are used as analytical lenses rather than prescriptive benchmarks to examine how existing approaches operationalise current standards, recommendations, and guidance.

\subsubsection{NIST PQC Standards}
\label{NISTPQC}

NIST established the PQC standardisation project \cite{nist_pqc} to develop and standardise PQC algorithms. Following a multi-round evaluation process, NIST published three PQC standards. \textit{FIPS 203: Module-Lattice-Based Key-Encapsulation Mechanism Standard (ML-KEM)} defines a key-encapsulation mechanism for secure shared-secret establishment \cite{nist_fips_203_2024}. \textit{FIPS 204: Module-Lattice-Based Digital Signature Standard (ML-DSA)} specifies a digital signature algorithm \cite{nist_fips_204_2024}, while \textit{FIPS 205: Stateless Hash-Based Digital Signature Standard (SLH-DSA)} defines a hash-based signature scheme \cite{nist_fips_205_2024}. We use the NIST PQC standards (FIPS 203--205) as the primary analytical baseline\footnote{ISO/IEC \cite{ISO_IEC_18033_2_Amd2_2026}, IETF \cite{rfc9958}, and ETSI \cite{ETSI_TR_103_618_2019} are treated as complementary references.} because they provide a finalised and bounded set of standardised PQC mechanisms with explicit specifications for consistent assessment. Alignment with these standards is therefore treated as an analytical dimension.

\subsubsection{PQC Migration Guidance}
\label{PQCMigGuide}

PQC migration extends beyond algorithm replacement and requires coordinated organisational and technical planning \cite{Campbell}. The PQCC migration roadmap \cite{pqcc_pqc_migration_roadmap_2025} provides a structured view of the migration process, covering \textit{(1) Preparation, (2) Baseline Understanding, (3) Planning and Execution, and (4) Monitoring and Evaluation}. These stages capture key activities for planning, executing, and sustaining PQC migration and are therefore used as an analytical lens to examine how existing approaches address its organisational and technical requirements of systematic PQC migration.

\subsubsection{Software Development Lifecycle (SDLC)}
\label{SDLC_ref}

PQC integration is not confined to cryptographic implementation; decisions affecting security may arise throughout software development, from requirements and design through coding, testing, deployment, and maintenance. This study therefore uses the SDLC, comprising planning, requirement analysis, design, coding, testing, deployment, and maintenance \cite{ibm_sdlc_2024}, to assess the SDLC coverage of existing approaches supporting PQC implementation and migration. 

\subsubsection{NIST Cybersecurity Framework Version 2.0}
\label{NISTCSFV2}

Secure PQC integration also involves cybersecurity activities that extend beyond the correct implementation of cryptographic primitives, including governance, risk identification, protection, detection, response, and recovery. NIST Cybersecurity Framework (CSF) 2.0 provides a widely established risk-based structure encompassing six core functions: \textit{Govern (GV), Identify (ID), Protect (PR), Detect (DE), Respond (RS), and Recover (RC)} \cite{nist_cswp_29_2024}. It is therefore used as a complementary analytical lens to examine whether existing PQC approaches address broader cybersecurity and organisational activities rather than focusing exclusively on cryptographic mechanisms.

\subsubsection{Deployment Maturity Level}
\label{DeplMat}

This SoK examines the deployment maturity of the approaches identified in the reviewed literature based on reported implementation evidence and validation activities, informed by the Technology Readiness Level (TRL) concept proposed by Mankins \cite{mankins1995technology}. The classification enables a consistent assessment of progression from theoretical concepts to practical deployment. Approaches are classified as \textit{Conceptual} when they constitute proposals or concepts without implementation; \textit{Prototype} when they provide proof-of-concept implementations demonstrating technical feasibility; \textit{Experimental} when implementations are validated through controlled experiments, benchmarks, simulations, or laboratory testing; \textit{Pilot} when evaluated in realistic or organisational environments using representative users, systems, or workloads; and \textit{Production-ready} when deployed with evidence of sustained real-world use.

\subsubsection{NIST Guidance for Testing and Evaluation}
\label{NISTTest}

Testing and evaluation are critical to ensuring secure PQC implementations, as implementation flaws may introduce vulnerabilities despite the use of standardised algorithms \cite{anonymous_pqc_sok_2026,toruan2026security}. This study assesses reported testing and evaluation evidence of the identified implementation and migration approaches using NIST IR 8397 \cite{nist_ir_8397_2021} and NIST PQC evaluation criteria \cite{nist_pqc_security_evaluation_criteria,nist_pqc_cost_evaluation_criteria} as reference points, covering software verification, implementation correctness, security, performance, and cost . Additional evidence, including interoperability testing, usability evaluation, and formal verification, is considered where reported by the reviewed studies.

\section{Methodology}
\label{Methodology}

\begin{figure*}[t]
\centering
\includegraphics[width=1\textwidth,height=0.5\textheight]{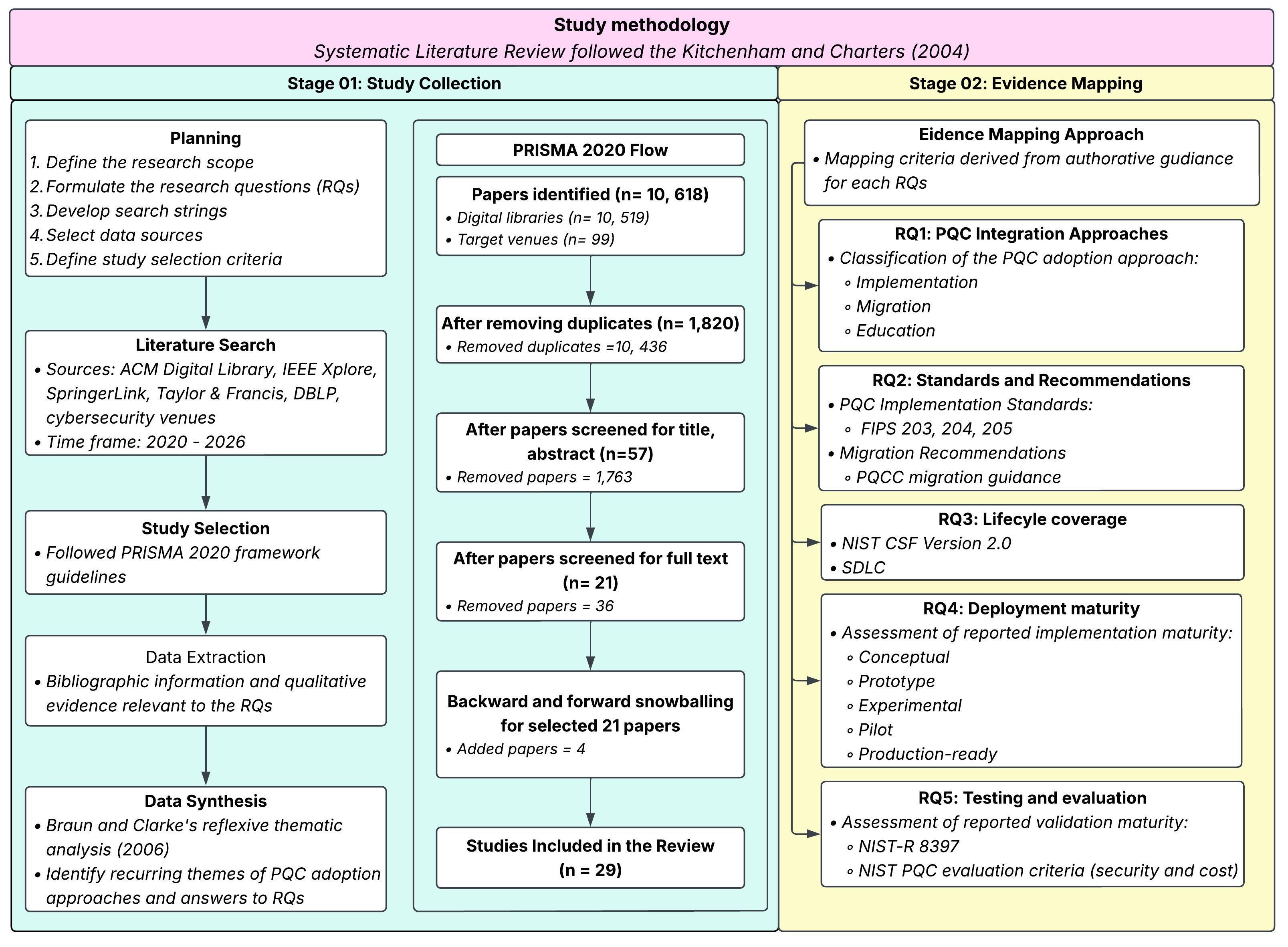}
\caption{Overview of the SoK methodology and evidence mapping workflow.}
\label{NIST_SoK_Method}
\end{figure*}

We followed a SoK methodology supported by the systematic literature review (SLR) process proposed by Kitchenham and Charters \cite{kitchenham_procedures_2004}. The methodology consists of two main stages: (i) study collection and (ii) evidence mapping, as illustrated in Figure~\ref{NIST_SoK_Method}. In the first stage, we identified and selected relevant literature using a structured review protocol. In the second stage, we synthesised and evaluated the extracted evidence against predefined dimensions derived from established standards, guidance, and frameworks.

\subsection{Study Collection}
\label{StudyCollection}

We developed a review protocol following Kitchenham and Charters \cite{kitchenham_procedures_2004} and defined the scope using PICOC framework. We operationalised the PICOC elements into search concepts and synonyms, which were combined using Boolean operators and adapted to the syntax of each database. We searched scholarly databases such as the ACM Digital Library, IEEE Xplore and selected cybersecurity venues for peer-reviewed English-language publications from January 2020 to July 2026 addressing PQC integration, implementation, migration, or developer education in software development (see Appendix~\ref{appendix:Methodology}). Study selection followed the PRISMA 2020 guidelines \cite{page_prisma_2021}. The search returned 10,519 records, with 99 additional records identified through targeted venue searches. After deduplication, title/abstract screening, full-text assessment against predefined criteria, and backward and forward snowballing, 25 studies were included. We used a predefined extraction form to record bibliographic information and evidence relevant to the research questions, including PQC standards, migration support, SDLC coverage, standards alignment, deployment, and evaluation evidence. We analysed the extracted evidence using Braun and Clarke's reflexive thematic analysis \cite{braun_using_2006}. Two authors independently reviewed the extracted evidence and assigned descriptive codes to segments addressing the research questions. Codes were compared across studies, grouped into candidate themes, and iteratively reviewed and refined against the underlying evidence. Themes were then defined, named, and mapped to the research questions to synthesise recurring patterns in PQC implementation, migration, and developer education. Coding discrepancies were discussed and resolved between the authors, and Cohen's kappa was used as a consistency check, yielding substantial agreement ($\kappa = 0.7945$).

\subsection{Evidence Mapping}
\label{StudyMapping}

We used evidence mapping to systematically characterise the approaches identified in the included studies. We derived mapping dimensions from authoritative standards and guidance, such as NIST FIPS 203-205, PQCC migration guidance, NIST CSF 2.0, and established SDLC practices. These sources served as analytical references for classification and comparison, rather than inclusion criteria. We operationalised each dimension into coding categories and applied the resulting scheme to the extracted evidence, iteratively refining category definitions to resolve ambiguities across studies. For example, the statement “... studied the performance of the algorithms...” from \cite{Ana_hekkala_implementing_2023} was coded as \textit{Performance evaluation} and mapped to RQ4 (Testing and evaluation).

Each study was coded and mapped across five dimensions:

\begin{enumerate}
\item \textbf{PQC Integration Approaches (RQ1):} Classification of the approach as supporting PQC implementation, migration, developer education, or a combination thereof.
\item \textbf{Standards and Migration Alignment (RQ2):} Identification of support for NIST PQC standards (FIPS 203-205) \cite{nist_fips_203_2024,nist_fips_204_2024,nist_fips_205_2024} and PQCC migration activities \cite{pqcc_pqc_migration_roadmap_2025}.
\item \textbf{SDLC and Cybersecurity Coverage (RQ3):} Assessment of SDLC activities \cite{ibm_sdlc_2024} and NIST CSF 2.0 functions \cite{nist_cswp_29_2024} addressed by each approach.
\item \textbf{Deployment Maturity (RQ4):} Classification from conceptual proposals to production deployments using the maturity levels defined in Section~\ref{DeplMat}.
\item \textbf{Evaluation Evidence (RQ5):} Assessment of reported testing and validation evidence using NIST IR 8397 \cite{nist_ir_8397_2021}, and NIST PQC evaluation criteria as reference points \cite{nist_pqc_security_evaluation_criteria, nist_pqc_cost_evaluation_criteria}, covering software verification, implementation correctness, security, performance, and cost. Additional evidence, including interoperability testing, usability evaluation, and formal verification, was recorded where reported.
\end{enumerate}

\subsection{Limitations of the Review}
\label{ReviewLimit}

Several limitations may affect the validity of this study. Relevant studies may have been missed because of database coverage, publication availability, or our search and selection criteria. Researchers may introduce subjectivity when interpreting qualitative data, coding studies, and mapping evidence. We mitigated these risks through predefined criteria and coding procedures, snowballing, independent coding by two authors, and discrepancy resolution through discussion. The review is further limited by its focus on peer-reviewed English-language literature, which may introduce publication and language bias. Finally, this SoK does not include expert or practitioner validation; consequently, the mapped approaches should not be interpreted as evidence of real-world deployment effectiveness or operational readiness.

\section{Results}
\label{Resul}
The analysis identified three categories of approaches supporting PQC integration in software development: \textit{implementation support}, \textit{migration support}, and \textit{developer education}. Implementation support and migration strategies address the technical and process-related activities involved in PQC integration, while developer education focuses on the development of knowledge and skills required for effective integration. 

\subsection{PQC Implementation Support}
\label{ResImpAp}

\begin{figure}[t]
\centering
\includegraphics[width=\columnwidth,keepaspectratio]{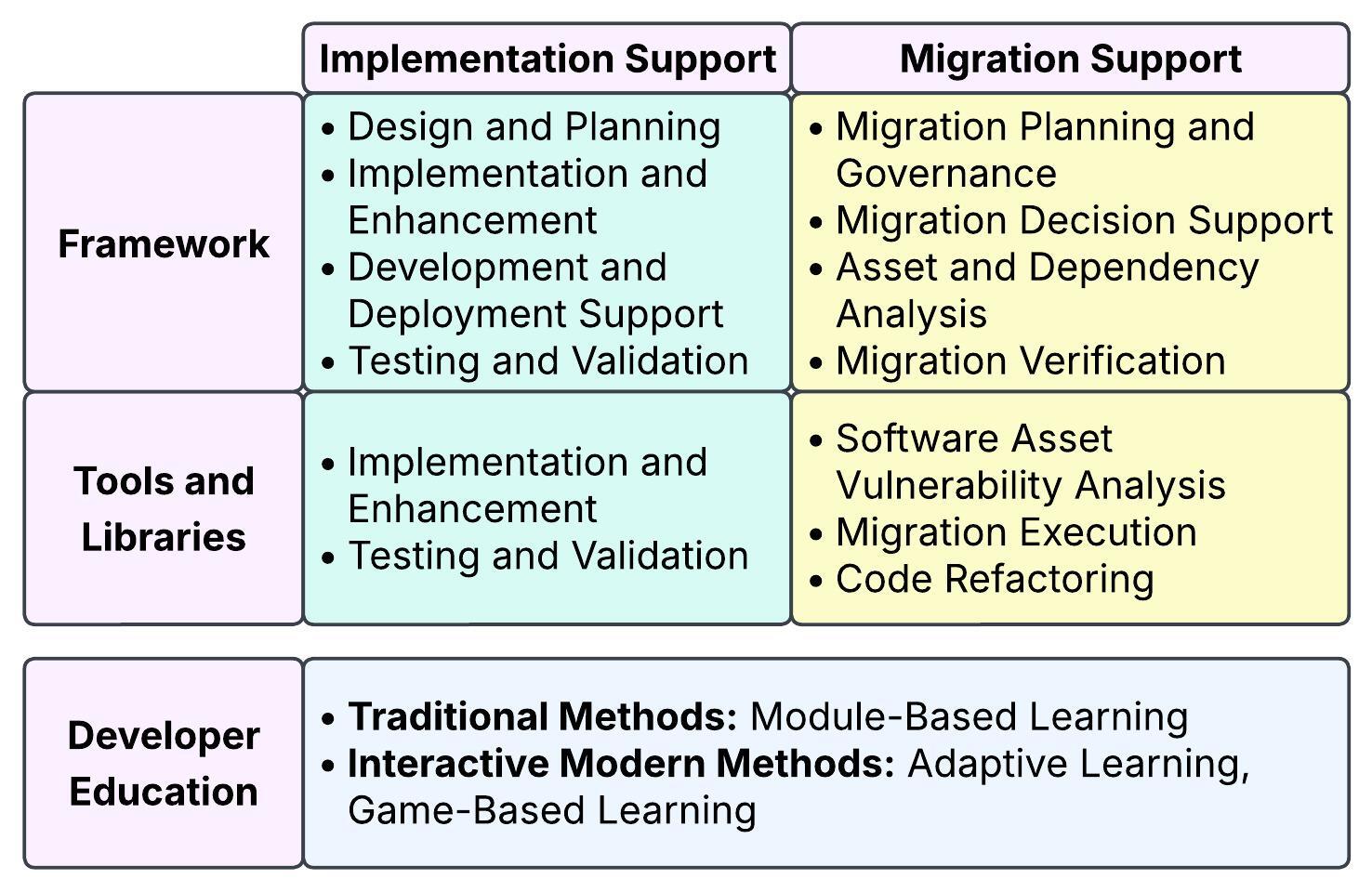}
\caption{Classification of approaches supporting PQC integration in software.}
\label{fig:approachTheme}
\end{figure}

\begin{table*}[ht]
\centering
\caption{Evidence-based assessment of PQC implementation support approaches across NIST PQC standards alignment, software development lifecycle (SDLC) coverage, NIST CSF v2 alignment, deployment maturity, and evaluation evidence.}
\label{tab:pqc_implementation_approaches}
\small
\resizebox{\textwidth}{!}{%
%\begin{tabular}{p{2cm} p{0.5cm} p{4.2cm} p{2cm} p{2cm} p{1.5cm} p{2cm} p{4cm}}
\begin{tabular}{
>{\raggedright\arraybackslash}m{2.1cm}
>{\centering\arraybackslash}m{0.5cm}
>{\raggedright\arraybackslash}m{4.7cm}
>{\centering\arraybackslash}m{1.5cm}
>{\centering\arraybackslash}m{2cm}
>{\centering\arraybackslash}m{1.5cm}
>{\centering\arraybackslash}m{2cm}
>{\raggedright\arraybackslash}m{4cm}
}
\hline
\makecell{\textbf{Approach}\\\textbf{Category}} &
\makecell{\textbf{Paper}\\\textbf{Ref.}} &
\makecell{\textbf{Used}\\\textbf{PQC Scheme}} &
\makecell{\textbf{NIST} \\\textbf{standards}\\\textbf{alignment}} &
\makecell{\textbf{Covered}\\\textbf{SDLC Stage}} &
\makecell{\textbf{Covered}\\\textbf{NIST CSF v2}} &
\makecell{\textbf{Deployment}\\\textbf{Maturity}} &
\makecell{\textbf{Reported}\\\textbf{Evaluation Evidence}} \\
\hline

\multicolumn{8}{c}{\textbf{\textit{Frameworks}}} \\
\hline

\multirow{2}{*}{\raggedright Design \& Planning} &
\cite{Ana_decisionAlgo} &
Scheme-agnostic &
S6 &
{\raggedright Design} &
GV &
Conceptual &
Performance evaluation \\

\cline{1-8}

\parbox{2cm}{\raggedright Implementation\\\& Enhancement} &
\cite{Ana_huelsing2021wireguard} &
Classic McEliece, SABER &
S3 &
Coding &
PR &
Experimental &
Formal security proof validation; Performance evaluation \\

\cline{1-8}

\multirow{2}{*}{\parbox{2cm}{\raggedright Development \& Deployment Support}} &
\cite{Ana_ojha2025adoption} &
Naoris core-based Dilithium &
S2 &
Coding &
PR &
Experimental &
Security testing (side-channel and cryptanalytic attacks) \\

&
\cite{Ana_tsili_scalable_2025} &
Falcon, Dilithium, SPHINCS+ &
S2, S4 &
Coding &
PR &
Experimental &
Performance evaluation \\

\cline{1-8}

\multirow{2}{*}{\parbox{2cm}{\raggedright Testing \& Validation}} &
\cite{Ana_kannwischer_improving_2022} &
Scheme-agnostic &
S6 &
Testing &
PR, DE &
Experimental &
Functional testing\\

&
\cite{Ana_e27020212} &
Kyber &
S2 &
Testing &
DE &
Experimental &
Protocol conformance; Performance evaluation \\

\hline

\multicolumn{8}{c}{\textbf{\textit{Tools and libraries}}} \\
\hline

\multirow{3}{*}{\parbox{2cm}{\raggedright Implementation \& Enhancement}} &
\cite{Ana_esser_sok_2024} &
Scheme-agnostic &
S6 &
Design &
PR &
Experimental &
Functional testing \\

&
\cite{Ana_hekkala_implementing_2023} &
CRYSTALS-Kyber, CRYSTALS-Dilithium, SABER, FrodoKEM &
S2 &
Coding &
PR &
Experimental &
KAT validation; Performance evaluation; Security testing (memory leak analysis) \\

&
\cite{Ana_saucedo-estrada_post-quantum_2020} &
NewHope &
S3 &
Coding &
PR &
Prototype &
Security testing (CCA evaluation) \\

\cline{1-8}

\multirow{2}{*}{\parbox{2cm}{\raggedright Testing \& Validation}} &
\cite{Ana_kannwischer_improving_2022} &
Scheme-agnostic &
S6 &
Testing &
PR, DE &
Experimental &
Functional testing \\

&
\cite{Ana_PQCLeo25} &
ML-DSA, CROSS-RSDPG, MAYO, SNOVA, Falcon, SPHINCS+, BIKE, FrodoKEM, ML-KEM, NTRU Prime, HQC &
S1, S3, S4, S5 &
Testing &
PR, DE &
Experimental &
Functional testing \\

\hline
\end{tabular}%
}

\begin{tablenotes}
\footnotesize
\item \textit{Abbreviations:} NIST CSF v2 functions: GV = Govern, PR = Protect, DE = Detect; KAT = Known Answer Test; CCA = Chosen-Ciphertext Attack; N/A = Not Applicable.

\item \textit{NIST PQC standards alignment coding:}
S1 = Standardised (final NIST-standardised schemes appeared in FIPS 203 - 205);
S2 = Pre-standardisation variant (earlier version of a scheme that subsequently became standardised);
S3 = NIST candidate (candidate evaluated in the NIST PQC process but not ultimately standardised); 
S4 = NIST-selected, not yet standardised (selected for future standardisation but not yet a final standard at the study/evaluation point); 
S5 = Non-NIST PQC (PQC scheme outside the NIST standardisation process); 
S6 = Scheme-agnostic (approach not tied to a specific PQC scheme). 

\end{tablenotes}

\end{table*}

PQC implementation support approaches provide the technical foundation for integrating PQC mechanisms into software systems \cite{anonymous_pqc_sok_2026}. As shown in Figure~\ref{fig:approachTheme}, the identified approaches include frameworks, tools, and libraries. Framework-based approaches support activities spanning design and planning \cite{Ana_decisionAlgo}, implementation and enhancement \cite{Ana_huelsing2021wireguard}, development and deployment support \cite{Ana_ojha2025adoption, Ana_tsili_scalable_2025}, and testing and validation \cite{Ana_kannwischer_improving_2022, Ana_e27020212}. Design and planning approaches support PQC algorithm and library selection \cite{Ana_decisionAlgo}, whereas implementation-oriented approaches integrate PQC mechanisms into existing software or protocols \cite{Ana_huelsing2021wireguard}. Development and deployment approaches support PQC implementation and operational integration, including certificate authorities for PQC-signed certificates \cite{Ana_ojha2025adoption, Ana_tsili_scalable_2025}. Testing and validation approaches evaluate PQC implementations and systems, including functional, conformance, security, and performance properties \cite{Ana_kannwischer_improving_2022, Ana_e27020212}. Tools and libraries similarly support direct PQC integration \cite{Ana_hekkala_implementing_2023, Ana_saucedo-estrada_post-quantum_2020}, hardness estimation for assessing the computational security of PQC schemes \cite{Ana_esser_sok_2024}, and testing of PQC schemes and implementations \cite{Ana_kannwischer_improving_2022, Ana_PQCLeo25}. 

Table~\ref{tab:pqc_implementation_approaches} summarises the evidence across NIST PQC standards alignment, SDLC coverage, NIST CSF, deployment maturity, and evaluation evidence.

\textbf{NIST PQC standards alignment: } The reviewed studies were classified by their relationship to the standardized NIST PQC schemes. One study implemented final standards, ML-KEM and ML-DSA (FIPS 203/204) \cite{Ana_PQCLeo25}. Four studies used pre-standardisation variants that contributed to NIST standards, including Kyber, Dilithium, Falcon, and SPHINCS+ \cite{Ana_ojha2025adoption, Ana_tsili_scalable_2025, Ana_e27020212, Ana_hekkala_implementing_2023}. Three studies implemented NIST candidates not ultimately standardised, including Classic McEliece, SABER, NewHope, FrodoKEM, MAYO, SNOVA, BIKE, and NTRU Prime \cite{Ana_huelsing2021wireguard, Ana_saucedo-estrada_post-quantum_2020, Ana_PQCLeo25}. Two considered schemes selected for future standardisation but not yet finalised at evaluation, namely Falcon, SPHINCS+, and HQC \cite{Ana_tsili_scalable_2025, Ana_PQCLeo25}. One employed the non-NIST scheme CROSS-RSDPG \cite{Ana_PQCLeo25}, while three studies adopted scheme-agnostic approaches \cite{Ana_decisionAlgo, Ana_kannwischer_improving_2022, Ana_esser_sok_2024}.

\textbf{SDLC coverage: } The identified approaches show limited and uneven coverage across the SDLC. Coding was addressed by five studies, primarily through cryptographic implementation of PQC schemes or libraries \cite{Ana_huelsing2021wireguard, Ana_ojha2025adoption, Ana_tsili_scalable_2025, Ana_hekkala_implementing_2023, Ana_saucedo-estrada_post-quantum_2020}. Testing was addressed by three studies through functional, conformance, and performance-oriented evaluation \cite{Ana_kannwischer_improving_2022, Ana_e27020212, Ana_PQCLeo25}. Two studies addressed design-oriented activities; one supported PQC algorithm and library selection \cite{Ana_decisionAlgo}, while another provided hardness estimation for assessing the computational security of PQC schemes \cite{Ana_esser_sok_2024}. No reviewed study explicitly addressed requirement analysis, deployment, or maintenance. 

\textbf{NIST CSF v2 alignment: } Alignment was concentrated on the Protect (PR) function, which concerns implementing safeguards to protect systems and data. Eight studies mapped to PR as their focus on PQC implementation, enhancement, development or testing \cite{Ana_huelsing2021wireguard, Ana_ojha2025adoption, Ana_tsili_scalable_2025, Ana_esser_sok_2024, Ana_hekkala_implementing_2023, Ana_saucedo-estrada_post-quantum_2020, Ana_kannwischer_improving_2022, Ana_PQCLeo25}. Detect (DE), which concerns identifying cybersecurity events or implementation problems, appeared in three studies through testing and validation \cite{Ana_kannwischer_improving_2022, Ana_e27020212, Ana_PQCLeo25}. Govern (GV), which concerns establishing and overseeing cybersecurity strategy and risk-management decisions, appeared only in the approach supporting PQC algorithm and library selection \cite{Ana_decisionAlgo}. No study addressed Identify (ID), concerning systematic identification of cryptographic assets and quantum-related risks, or Respond (RS) and Recover (RC), concerning incident management and restoration of affected operations.

\textbf{Deployment maturity:} Deployment evidence was concentrated at the experimental level, with eight studies validating their approaches through controlled experiments, benchmarks, or laboratory assessment \cite{Ana_huelsing2021wireguard, Ana_ojha2025adoption, Ana_tsili_scalable_2025, Ana_kannwischer_improving_2022, Ana_e27020212, Ana_esser_sok_2024, Ana_hekkala_implementing_2023, Ana_PQCLeo25}. One study was conceptual, presenting design and decision support without implementation \cite{Ana_decisionAlgo}, and one provided a prototype to demonstrate technical feasibility \cite{Ana_saucedo-estrada_post-quantum_2020}. No study demonstrated pilot deployment in a realistic organisational setting or production deployment with sustained real-world use.

\textbf{Evaluation evidence:} Evaluation primarily addressed technical correctness, security, and performance. Performance was assessed through measures such as latency \cite{Ana_decisionAlgo, Ana_huelsing2021wireguard, Ana_tsili_scalable_2025, Ana_e27020212, Ana_hekkala_implementing_2023, Ana_PQCLeo25}, while security evaluation included formal proofs, side-channel and cryptanalytic testing, memory-leak analysis, and CCA-based evaluation \cite{Ana_huelsing2021wireguard, Ana_ojha2025adoption, Ana_hekkala_implementing_2023, Ana_saucedo-estrada_post-quantum_2020}. Implementation correctness was assessed through KAT and protocol-conformance testing \cite{Ana_hekkala_implementing_2023, Ana_e27020212}, and functional testing was reported in several studies \cite{Ana_kannwischer_improving_2022, Ana_esser_sok_2024, Ana_PQCLeo25}. However, evaluation of interoperability, usability, cost, and deployment testing remained limited or absent.

\subsection{PQC Migration Support}
\label{ResMigAp}

\begin{table*}[ht]
\centering
\caption{Evidence-based assessment of PQC migration approaches across PQCC migration activity coverage, software development lifecycle (SDLC) coverage, NIST CSF v2 alignment, deployment maturity, and evaluation evidence.}
\label{tab:pqc_migration_classification}
\small
\resizebox{\textwidth}{!}{%
\begin{tabular}{
>{\raggedright\arraybackslash}m{3cm}
>{\centering\arraybackslash}m{0.7cm}
>{\centering\arraybackslash}m{1.8cm}
>{\raggedright\arraybackslash}m{5.2cm}
>{\centering\arraybackslash}m{1.5cm}
>{\centering\arraybackslash}m{2cm}
>{\raggedright\arraybackslash}m{2cm}
}
\hline
\makecell{\textbf{Approach}\\\textbf{Category}} &
\makecell{\textbf{Paper}\\\textbf{Ref.}} &
\makecell{\textbf{PQC}\\\textbf{Migration}\\\textbf{Activity}} &
\makecell{\textbf{Covered}\\\textbf{SDLC Stage}} &
\makecell{\textbf{Covered}\\\textbf{NIST}\\\textbf{CSF v2}} &
\makecell{\textbf{Deployment}\\\textbf{Maturity}} &
\makecell{\textbf{Reported}\\\textbf{Evaluation}\\\textbf{Evidence}} \\
\hline

\multicolumn{7}{c}{\textbf{\textit{Frameworks}}} \\
\hline

\multirow{4}{*}{\parbox{3cm}{\raggedright Migration Planning\\and Governance}} &
\cite{Ana_MigFrDepen} &
1,2,3,4 &
Planning, Requirement analysis, Design, Coding, Testing, Maintenance &
GV, ID, PR, DE, RS, RC &
Conceptual &
Not reported \\

&
\cite{Ana_MigFrameOrg} &
1,2,3,4 &
Planning, Requirement analysis, Coding, Testing, Deployment, Maintenance &
GV, ID, PR, DE &
Prototype &
Scenario-based evaluation \\

&
\cite{Ana_PMMP} &
1 &
Planning, Requirement analysis, Coding, Testing &
GV, ID, DE &
Conceptual &
Theoretical validation \\

&
\cite{Ana_FinanceMig} &
1,2,3,4 &
Planning, Requirement analysis, Testing, Deployment &
GV, ID, PR, DE &
Conceptual &
Not reported \\

\cline{1-7}

\parbox{3cm}{\raggedright Migration Decision\\Support} &
\cite{Ana_decisionAlgo} &
3 &
Design &
GV &
Conceptual &
Performance evaluation \\

\cline{1-7}

\parbox{3cm}{\raggedright Asset and Dependency\\Analysis} &
\cite{Ana_MigFrDepen} &
2 &
Planning, Requirement analysis &
ID &
Conceptual &
No reported \\

\cline{1-7}

\parbox{3cm}{\raggedright Migration Verification} &
\cite{Ana_e27020212} &
4 &
Testing &
DE &
Experimental &
Protocol conformance; Performance evaluation \\

\hline

\multicolumn{7}{c}{\textbf{\textit{Tools and libraries}}} \\
\hline

\parbox{3cm}{\raggedright Software Asset\\Vulnerability Analysis} &
\cite{Ana_rattanavipanon_toolchain_2025} &
3 &
Requirement analysis, Testing &
ID, DE &
Experimental &
Functional testing \\

\cline{1-7}

\parbox{3cm}{\raggedright Migration\\Execution} &
\cite{Ana_android_Migr} &
3 &
Coding &
PR &
Experimental &
Performance evaluation \\

\cline{1-7}

\multirow{2}{*}{\parbox{3cm}{\raggedright Code Refactoring}} &
\cite{Ana_ccPASTpqc} &
3 &
Coding &
ID, PR &
Prototype &
Functional testing \\

&
\cite{Ana_legacyToQuantumConvert} &
2,3 &
Planning, Requirement analysis, Coding &
PR, DE &
Conceptual &
Not reported \\

\hline
\end{tabular}%
}

\begin{tablenotes}
\footnotesize
\item \textit{Abbreviations:} PQC migration activity categories: (1) Preparation; (2) Baseline Understanding; (3) Planning and Execution; and (4) Monitoring and Evaluation; NIST CSF v2 functions: GV = Govern, ID = Identify, PR = Protect, DE = Detect, RS = Respond, RC = Recover.
\end{tablenotes}

\end{table*}

PQC migration approaches support the transition from classical cryptography to quantum-resistant alternatives through migration planning, asset and dependency analysis, decision-making, implementation, refactoring, and verification \cite{naether_migrating_2024, migration_Comparison}. As shown in Figure~\ref{fig:approachTheme}, the ten studies comprise two categories: frameworks and tools or libraries. Frameworks primarily address migration planning, governance, and lifecycle management \cite{Ana_MigFrDepen, Ana_MigFrameOrg, Ana_PMMP, Ana_FinanceMig}, with additional support for decision-making \cite{Ana_decisionAlgo}, asset and dependency analysis \cite{Ana_MigFrDepen}, and verification \cite{Ana_e27020212}. Tools and libraries focus on technical activities, including asset vulnerability analysis \cite{Ana_rattanavipanon_toolchain_2025}, migration execution \cite{Ana_android_Migr}, and code refactoring \cite{Ana_ccPASTpqc, Ana_legacyToQuantumConvert}. Overall, existing PQC migration approaches address key activities across the migration lifecycle, but remain fragmented, with limited end-to-end integration.

Table~\ref{tab:pqc_migration_classification} summarises the identified migration approaches and their evidence-based assessment across PQC migration activity coverage, SDLC integration, deployment maturity, and evaluation evidence.

\textbf{PQCC migration activity coverage:} The migration approaches were classified across the four PQCC roadmap stages: Preparation, Baseline Understanding, Planning and Execution, and Monitoring and Evaluation. Planning and execution received the strongest coverage, with eight studies addressing migration planning, decision-making for selecting algorithms or libraries, execution, or vulnerability analysis \cite{Ana_MigFrDepen, Ana_MigFrameOrg, Ana_FinanceMig, Ana_decisionAlgo, Ana_rattanavipanon_toolchain_2025, Ana_android_Migr, Ana_ccPASTpqc, Ana_legacyToQuantumConvert}. 
Preparation was addressed by four studies, primarily through migration planning and governance activities \cite{Ana_MigFrDepen, Ana_MigFrameOrg, Ana_PMMP, Ana_FinanceMig}, while baseline understanding was addressed by four studies through cryptographic asset and dependency analysis and legacy-code assessment \cite{Ana_MigFrDepen, Ana_MigFrameOrg, Ana_FinanceMig, Ana_legacyToQuantumConvert}. Monitoring and evaluation were addressed, with four studies covering verification, testing, or migration evaluation \cite{Ana_MigFrDepen, Ana_MigFrameOrg, Ana_FinanceMig, Ana_e27020212}. Notably, three migration-planning frameworks \cite{Ana_MigFrDepen, Ana_MigFrameOrg, Ana_FinanceMig} covered all four stages, providing end-to-end lifecycle coverage.

\textbf{SDLC coverage: } The approaches provide broad but uneven SDLC coverage, with greater emphasis on planning and implementation than the complete lifecycle. Requirement analysis was most frequently addressed, with six studies covering cryptographic asset and dependency identification, vulnerability assessment, migration strategy selection, and requirements definition \cite{Ana_MigFrDepen, Ana_MigFrameOrg, Ana_PMMP, Ana_FinanceMig, Ana_rattanavipanon_toolchain_2025, Ana_legacyToQuantumConvert}. Planning was addressed by five studies through migration planning and governance activities \cite{Ana_MigFrDepen, Ana_MigFrameOrg, Ana_PMMP, Ana_FinanceMig, Ana_legacyToQuantumConvert}, while coding was addressed by six studies through migration execution and code-refactoring activities \cite{Ana_MigFrDepen, Ana_MigFrameOrg, Ana_PMMP, Ana_android_Migr, Ana_ccPASTpqc, Ana_legacyToQuantumConvert}. Testing was covered by six studies through verification, functional testing, and evaluation \cite{Ana_MigFrDepen, Ana_MigFrameOrg, Ana_PMMP, Ana_FinanceMig, Ana_e27020212, Ana_rattanavipanon_toolchain_2025}. Design \cite{Ana_MigFrDepen, Ana_decisionAlgo}, deployment \cite{Ana_MigFrameOrg, Ana_FinanceMig}, and maintenance \cite{Ana_MigFrDepen, Ana_MigFrameOrg} received limited coverage, with only two studies addressing each stage. 

\textbf{NIST CSF v2 alignment:} Alignment spanned all six cybersecurity activities, with the strongest coverage of Detect (DE). Govern (GV), concerning cybersecurity strategy, policy, and risk-management decisions, was addressed by five studies through planning, governance, financial planning, algorithm selection, and verification \cite{Ana_MigFrDepen, Ana_MigFrameOrg, Ana_PMMP, Ana_FinanceMig, Ana_decisionAlgo}. Identify (ID), concerning organisational assets, dependencies, and cybersecurity risks, was also addressed by six studies through asset and dependency analysis, and vulnerability assessment \cite{Ana_MigFrDepen, Ana_MigFrameOrg, Ana_PMMP, Ana_FinanceMig, Ana_rattanavipanon_toolchain_2025, Ana_ccPASTpqc}. Protect (PR), which concerns implementing safeguards to protect systems and data, was represented in six studies through PQC migration execution and code-refactoring activities \cite{Ana_MigFrDepen, Ana_MigFrameOrg, Ana_FinanceMig, Ana_ccPASTpqc, Ana_legacyToQuantumConvert, Ana_android_Migr}. Detect (DE), which concerns identifying cybersecurity events and vulnerabilities, was addressed by seven studies through vulnerability analysis, testing, verification, and migration assessment activities \cite{Ana_MigFrDepen, Ana_MigFrameOrg, Ana_PMMP, Ana_FinanceMig, Ana_rattanavipanon_toolchain_2025, Ana_e27020212, Ana_legacyToQuantumConvert}. Respond (RS) and Recover (RC) received minimal coverage, with only one study mapped to these functions \cite{Ana_MigFrDepen}.

\textbf{Deployment maturity: } PQC migration approaches remain concentrated at conceptual, prototype, and experimental levels, with no evidence of production-ready deployment. Five approaches were conceptual, providing proposed frameworks or migration support without sufficient implementation evidence \cite{Ana_MigFrDepen, Ana_PMMP, Ana_FinanceMig, Ana_decisionAlgo, Ana_legacyToQuantumConvert}. Two were prototypes demonstrating technical feasibility through proof-of-concept implementations \cite{Ana_MigFrameOrg, Ana_ccPASTpqc}, and three were experimental, evaluated through controlled testing such as performance evaluation \cite{Ana_e27020212, Ana_rattanavipanon_toolchain_2025, Ana_android_Migr}. No approach demonstrated pilot or production deployment with sustained evaluation in realistic environments.

\textbf{Evaluation evidence: } Evaluation evidence was limited and primarily technical, focusing on functional correctness, protocol conformance, and performance. Functional testing was reported for vulnerability analysis and code-refactoring approaches \cite{Ana_rattanavipanon_toolchain_2025, Ana_ccPASTpqc}, while migration verification included protocol-conformance and performance evaluation \cite{Ana_e27020212}. Performance was also assessed in migration decision-support and implementation approaches \cite{Ana_decisionAlgo, Ana_android_Migr}. One study used scenario-based evaluation \cite{Ana_MigFrameOrg}, another provided theoretical validation \cite{Ana_PMMP}, and three reported no evaluation evidence \cite{Ana_MigFrDepen, Ana_FinanceMig, Ana_legacyToQuantumConvert}.

\subsection{Developer Education}
\label{ResEduAp}

\begin{table*}[ht]
\centering
\caption{Evidence-based assessment of developer education approaches across knowledge coverage, NIST CSF v2 alignment, learning modality, and evaluation evidence.}
\label{tab:pqc_education_classification}
\small
\resizebox{\textwidth}{!}{%
\begin{tabular}{
>{\raggedright\arraybackslash}m{1.5cm}
>{\centering\arraybackslash}m{0.7cm}
>{\raggedright\arraybackslash}m{5.8cm}
>{\centering\arraybackslash}m{1.5cm}
>{\centering\arraybackslash}m{2cm}
>{\raggedright\arraybackslash}m{3.4cm}
}
\hline
\makecell{\textbf{Approach}\\\textbf{Category}} &
\makecell{\textbf{Paper}\\\textbf{Ref.}} &
\makecell{\textbf{PQC Knowledge}\\\textbf{Coverage}} &
\makecell{\textbf{Covered}\\\textbf{NIST CSF v2}} &
\makecell{\textbf{Learning}\\\textbf{Modality}} &
\makecell{\textbf{Reported}\\\textbf{Evaluation Evidence}} \\
\hline

\multicolumn{6}{c}{\textbf{\textit{Traditional Methods}}} \\
\hline

\multirow{5}{*}{\parbox{2.5cm}{\raggedright Module-Based\\Learning}} &
\cite{Ana_borrelli_towards_2025} &
CRYSTALS-Kyber, CRYSTALS-Dilithium, SPHINCS+, FALCON &
PR.AT &
In-person instruction &
User study (Academic learners) \\

&
\cite{Ana_borrelli_designing_2024} &
NIST PQC standardisation progress &
PR.AT &
In-person instruction &
User study (Academic learners) \\

&
\cite{Ana_edComScSociety} &
PQC protocols and NIST standards, quantum risk and threat analysis &
PR.AT &
In-person instruction &
User study (Academic learners) \\

&
\cite{Ana_PQCIP} &
Quantum vulnerability assessment, NIST PQC standards, PQC migration &
PR.AT &
Online instruction &
User study (IT professionals) \\

&
\cite{Ana_EducModule} &
NIST PQC standardisation &
PR.AT &
Conceptual design &
Not reported \\

\hline

\multicolumn{6}{c}{\textbf{\textit{Interactive Modern Methods}}} \\
\hline

\parbox{2.5cm}{\raggedright Adaptive\\Learning} &
\cite{Ana_PQCLabEdu} &
Lattice parameter selection &
PR.AT &
Interactive prototype &
Functional testing of prototype; User study (Academic learners) \\

\cline{1-6}

\parbox{2.5cm}{\raggedright Game-Based\\Learning} &
\cite{Ana_abdelhamid_wip_2024} &
Basic PQC concepts &
PR.AT &
Interactive prototype &
User study (Academic learners) \\

\hline
\end{tabular}%
}

\begin{tablenotes}
\footnotesize
\item \textit{Abbreviations:} NIST CSF v2 function: PR.AT = Protect (Awareness and Training).
\item \textit{Evaluation evidence primarily concerns immediate learning experiences or short-term outcomes; longer-term knowledge retention, transfer to PQC implementation tasks, and sustained changes in secure development behaviour were rarely reported.}
\end{tablenotes}

\end{table*}

Developer education approaches support the knowledge and skills required for PQC implementation and migration, which supports overall PQC integration \cite{Ana_borrelli_designing_2024, Ana_borrelli_towards_2025}. As shown in Figure~\ref{fig:approachTheme}, the identified approaches comprise traditional and interactive modern methods. Traditional approaches primarily use module-based learning through curricula, educational modules, and formal training \cite{Ana_borrelli_towards_2025, Ana_borrelli_designing_2024, Ana_edComScSociety, Ana_PQCIP, Ana_EducModule}. Interactive methods include adaptive learning and game-based learning. Adaptive learning approaches provide personalised learning environments through interactive tools that adjust learning content according to learner requirements \cite{Ana_PQCLabEdu}, while game-based approaches employ game mechanics and practical exercises to support PQC learning and learner engagement \cite{Ana_abdelhamid_wip_2024}. 

Table~\ref{tab:pqc_education_classification} summarises the identified developer education approaches across PQC knowledge coverage, NIST CSF v2 alignment, learning modality, and evaluation evidence.

\textbf{PQC knowledge coverage: } The reviewed approaches primarily provide conceptual knowledge of PQC schemes, standards, protocols, and quantum threats, with limited practical implementation skills. Four module-based approaches cover NIST PQC algorithms, standardisation, protocols, and quantum-related risks \cite{Ana_borrelli_towards_2025, Ana_borrelli_designing_2024, Ana_edComScSociety, Ana_PQCIP, Ana_EducModule}. The adaptive learning approach addresses lattice parameter selection \cite{Ana_PQCLabEdu}, whereas the game-based approach covers basic PQC concepts \cite{Ana_abdelhamid_wip_2024}. Notably, only one study combines PQC knowledge with quantum-vulnerability assessment and PQC migration, providing broader coverage relevant to software practitioners, although practical implementation content remains limited \cite{Ana_PQCIP}. Overall, education approaches establish foundational knowledge but provide limited practical skills for secure implementation, testing, deployment, and migration. 

\textbf{NIST CSF v2 alignment: } Alignment was uniformly limited to Protect (PR), specifically the awareness and training (PR.AT) category. All seven reviewed education approaches were mapped to PR.AT, reflecting their focus on developing learners' knowledge and awareness of PQC concepts, standards, quantum-related threats, and migration considerations \cite{Ana_borrelli_towards_2025, Ana_borrelli_designing_2024, Ana_edComScSociety, Ana_PQCIP, Ana_EducModule, Ana_PQCLabEdu, Ana_abdelhamid_wip_2024}. Two studies additionally covered quantum threat/risk analysis, vulnerability assessment, and PQC migration \cite{Ana_edComScSociety, Ana_PQCIP}; although relevant to Identify (ID), these did not constitute explicit CSF ID alignment. No approach addressed Detect (DE), Respond (RS), or Recover (RC), including implementation-problem detection, incident management, or operational recovery.

\textbf{Learning modality: } The reviewed developer education approaches employ three main learning modalities: in-person instruction, online instruction, and interactive prototypes. In-person instruction is the most common modality, used by three module-based approaches \cite{Ana_borrelli_towards_2025, Ana_borrelli_designing_2024, Ana_edComScSociety}. One module-based approach uses online instruction \cite{Ana_PQCIP}. Interactive prototypes are used by the adaptive and game-based approaches to provide an interactive learning experience \cite{Ana_PQCLabEdu, Ana_abdelhamid_wip_2024}. One approach remains at the conceptual design stage and therefore does not report an implemented learning modality \cite{Ana_EducModule}. 

\textbf{Evaluation evidence: } Evaluation evidence was reported for six of the seven reviewed education approaches, mostly through user studies. Five approaches involved academic learners \cite{Ana_borrelli_towards_2025, Ana_borrelli_designing_2024, Ana_edComScSociety, Ana_PQCLabEdu, Ana_abdelhamid_wip_2024}, while one involved IT professionals \cite{Ana_PQCIP}. The adaptive learning approach additionally combined functional testing of its prototype with a user study \cite{Ana_PQCLabEdu}, whereas the game-based approach was evaluated through a user study with academic learners \cite{Ana_abdelhamid_wip_2024}. In contrast, the conceptual module design reported no evaluation evidence \cite{Ana_EducModule}. Overall, the reported evaluations primarily concern immediate learning experiences or short-term outcomes, with limited evidence of longer-term knowledge retention, transfer to PQC implementation tasks, or sustained changes in secure development behaviour.

\section{Discussion}
\label{Discuss}

The findings show that existing PQC approaches provide important technical and educational foundations but remain fragmented across the software development lifecycle. Implementation, migration, and education approaches address distinct aspects of PQC integration, yet rarely form an integrated pathway from requirements analysis to maintenance activities. 

\subsection{PQC Implementation Support}
\label{DisImReady}

The evidence indicates that the PQC implementation landscape has progressed from conceptual proposals toward technically demonstrable implementation mechanisms \cite{Ana_huelsing2021wireguard, Ana_ojha2025adoption, Ana_tsili_scalable_2025, Ana_hekkala_implementing_2023}, but this progress has not yet translated into comprehensive support for secure PQC integration. The main issue is therefore not the absence of implementation support, but the fragmentation and limited maturity of the available support. Existing approaches provide useful capabilities for individual activities, particularly cryptographic implementation and technical validation \cite{Ana_hekkala_implementing_2023, Ana_saucedo-estrada_post-quantum_2020, Ana_kannwischer_improving_2022, Ana_e27020212, Ana_PQCLeo25}, but provide limited evidence of coordinated support across the SDLC and organisational context.

The emphasis on coding and testing indicates that PQC integration is treated primarily as an implementation problem rather than a lifecycle-wide engineering activity. Secure integration also requires requirements analysis, cryptographic asset and dependency identification, deployment, monitoring, and maintenance. Without support for these activities, existing approaches may help engineers implement or validate a selected mechanism without addressing the decisions needed for sustainable integration. Thus, the current landscape provides implementation components rather than an integrated lifecycle process. This fragmentation also creates coordination challenges because PQC activities are interdependent; for example, asset identification informs algorithm selection, which affects implementation, testing, and deployment \cite{anonymous_pqc_sok_2026}. Consequently, although individual support exists, evidence of their systematic coordination across the SDLC remains limited \cite{Ana_decisionAlgo, Ana_huelsing2021wireguard, Ana_ojha2025adoption, Ana_tsili_scalable_2025, Ana_kannwischer_improving_2022, Ana_e27020212}.

The strong concentration on the NIST CSF v2 Protect function indicates that PQC implementation is largely framed as strengthening cryptographic protection. While this focus is understandable given the technical nature of PQC, it provides an incomplete view of the broader implementation problem. However, secure PQC implementation also requires identifying vulnerable assets, assessing risks and dependencies, defining responsibilities, and managing transition-related issues. Limited coverage of Identify and Govern therefore reveals gaps in asset discovery, risk management, policy, and organisational coordination. Thus, current research tends to narrow PQC implementation to cryptographic protection rather than treating it as a long-term organisational cybersecurity transition \cite{Ana_decisionAlgo}.

The limited coverage of final NIST-standardised mechanisms constrains the applicability of existing findings. Much of the literature evaluates earlier algorithm variants or NIST candidates \cite{Ana_ojha2025adoption, Ana_tsili_scalable_2025, Ana_e27020212, Ana_hekkala_implementing_2023, Ana_huelsing2021wireguard, Ana_saucedo-estrada_post-quantum_2020}. These studies are valuable for establishing implementation and performance knowledge, these findings cannot automatically represent the behaviour, interfaces, configurations, interoperability, or engineering requirements of final standards. This temporal gap is particularly important as PQC integration shifts from evaluating candidate algorithms to securely integrating standardised mechanisms into software systems. Future research should therefore focus on standards-conformant implementations and their practical integration, including API design, configuration, interoperability, dependency management, testing, and migration, particularly following the finalised PQC standards such as FIPS 203 - 205 \cite{nist_fips_203_2024,nist_fips_204_2024,nist_fips_205_2024}.

The predominance of experimental evidence shows that PQC mechanisms can be implemented and evaluated under controlled conditions, but not that they can be reliably integrated and sustained in realistic environments. Operational deployment introduces constraints such as legacy dependencies, heterogeneous infrastructure, interoperability, organisational processes, developer expertise, configuration, and maintenance that laboratory studies may not capture. The absence of pilot and production evidence therefore represents a gap between technical feasibility and reliable deployment. This gap is particularly important for PQC, given transitional states, evolving standards, dependencies, and repeated cryptographic changes. Evidence from realistic settings is therefore needed to determine whether current approaches remain effective under these conditions.

Existing studies evaluate performance, functional correctness, protocol conformance, and selected security properties, including side-channel resistance \cite{Ana_decisionAlgo, Ana_huelsing2021wireguard, Ana_tsili_scalable_2025, Ana_e27020212, Ana_hekkala_implementing_2023, Ana_PQCLeo25}. However, these evaluations provide limited evidence of whether developers can use and maintain such mechanisms securely in practice, with less attention to API usability, developer understanding, configuration errors, interoperability, maintainability, migration effort, and organisational cost. This is important because cryptographic security depends not only on the underlying algorithm but also on how developers select, configure, invoke, test, and maintain them. Poorly designed interfaces or unclear configuration requirements can introduce practical weaknesses despite technical correctness \cite{toruan2026security}. Evaluation also shows limited systematic alignment with generic evaluation guidance such as NIST IR 8397, including fuzzing and systematic verification \cite{nist_ir_8397_2021}. Overall, the literature demonstrates technical feasibility more strongly than secure and sustainable PQC integration by software engineers and organisations.

\begin{tcolorbox}[
title=\textbf{Takeaway},
colback=gray!5,
colframe=black,
boxrule=0.5pt,
arc=0.5mm,
left=0.5mm,
right=0.5mm,
top=0.5mm,
bottom=0.5mm
]
Existing PQC implementation support provides technical building blocks rather than integrated lifecycle support. The key gap is the lack of empirically validated approaches that connect algorithm and library selection, implementation, testing, deployment, and maintenance while coordinating the required technical, developer, and organisational activities.
\end{tcolorbox}

\subsection{PQC Migration Support}
\label{DisMigReady}

The evidence indicates that the PQC migration landscape has developed a range of approaches supporting migration planning to migration verification \cite{Ana_MigFrDepen, Ana_MigFrameOrg, Ana_PMMP, Ana_FinanceMig, Ana_decisionAlgo, Ana_rattanavipanon_toolchain_2025, Ana_android_Migr, Ana_ccPASTpqc, Ana_legacyToQuantumConvert}. However, these approaches remain fragmented across frameworks, tools, and libraries rather than being integrated into a coherent migration process. Framework-based approaches primarily support migration planning, governance, and decision-making \cite{Ana_MigFrDepen, Ana_MigFrameOrg, Ana_PMMP, Ana_FinanceMig, Ana_decisionAlgo}, whereas tools and libraries mainly address specific technical activities such as vulnerability analysis and code refactoring \cite{Ana_rattanavipanon_toolchain_2025, Ana_android_Migr, Ana_ccPASTpqc, Ana_legacyToQuantumConvert}. Thus, the principal limitation is not the absence of support for individual migration activities, but the fragmentation and limited integration of these supporting approaches across the migration lifecycle.

The concentration of existing approaches on planning, requirement analysis, coding, and testing suggests that PQC migration is primarily treated as a transition and implementation problem rather than as a continuous software development lifecycle activity. Although these activities are necessary, sustainable migration also requires coordinated support for deployment, maintenance, and monitoring. For example, cryptographic asset identification should inform migration priorities; migration requirements should guide algorithm selection and implementation; implementation changes should be verified before deployment and so on \cite{naether_migrating_2024, migration_Comparison}. Existing approaches provide support for several of these activities, but limited evidence demonstrates how they are systematically connected. This fragmentation is particularly visible in the distinction between frameworks that define migration processes and tools that execute individual technical tasks. Consequently, current approaches provide migration capabilities rather than an integrated migration pathway, leaving limited evidence of how organisations and developers can coordinate migration activities from initial preparation through execution and continued operation.

The NIST CSF v2 mapping provides further evidence of this imbalance. The reviewed approaches show comparatively strong representation of Govern, Identify, Protect, and Detect, while Respond and Recover receive minimal attention. This distribution suggests that PQC migration research has primarily focused on establishing migration governance, understanding cryptographic assets and risks, implementing replacement mechanisms, and identifying migration problems \cite{Ana_MigFrDepen, Ana_MigFrameOrg, Ana_PMMP, Ana_FinanceMig, Ana_rattanavipanon_toolchain_2025, Ana_e27020212}. These functions are essential, but migration can also introduce operational and security problems that require response and recovery. For example, incorrect migration decisions, configuration errors, or unexpected dependencies may require organisations to restore affected services and revise migration plans. The limited representation of Respond and Recover therefore indicates that existing approaches provide stronger support for preparing for and executing migration than for managing failures and recovering from migration-related disruption. This suggests a need to conceptualise PQC migration not only as a cryptographic transition but also as a broader cybersecurity and resilience process.

The uneven coverage of the PQCC migration activities \cite{pqcc_pqc_migration_roadmap_2025} reinforces this observation. Most approaches concentrate on Planning and Execution, while comparatively fewer address Monitoring and Evaluation. Several frameworks span preparation, baseline understanding, planning and execution, and monitoring and evaluation \cite{Ana_MigFrDepen, Ana_MigFrameOrg, Ana_FinanceMig}; however, broad activity coverage does not necessarily demonstrate operational integration or continuous use. Monitoring is particularly important because PQC migration occurs within an evolving cryptographic environment in which algorithms, dependencies, configurations, and organisational requirements can change over time \cite{naether_migrating_2024, migration_Comparison}. A migration process that ends after implementation and initial verification may therefore provide only temporary assurance. The limited attention to continuous monitoring and evaluation suggests that current migration research more strongly addresses how to initiate and perform migration than how to manage and sustain the migrated state.

Furthermore, limited deployment maturity constrains evidence of real-world migration. Most approaches remain at conceptual, prototype, or experimental levels, with no identified evidence of pilot- or production-level deployment. Conceptual frameworks provide useful guidance for structuring migration activities, while prototypes and experimental tools demonstrate technical feasibility \cite{Ana_MigFrDepen, Ana_MigFrameOrg, Ana_PMMP, Ana_FinanceMig, Ana_ccPASTpqc, Ana_e27020212, Ana_rattanavipanon_toolchain_2025, Ana_android_Migr}. However, migration in operational software environments introduces constraints that controlled evaluations may not capture, including legacy dependencies, undocumented cryptographic usage, interoperability requirements, organisational responsibilities \cite{naether_migrating_2024, migration_Comparison}. The absence of pilot and production evidence therefore represents more than a maturity gap; it reflects a missing stage of empirical knowledge between demonstrating that migration mechanisms are technically feasible and establishing that they can be applied reliably within realistic organisational environments.

Existing evaluations also provide stronger evidence of technical functionality than of migration effectiveness in practice. Reported evidence includes functional testing, protocol conformance, performance evaluation, scenario-based evaluation, and theoretical validation \cite{Ana_rattanavipanon_toolchain_2025, Ana_ccPASTpqc, Ana_e27020212, Ana_decisionAlgo, Ana_android_Migr, Ana_MigFrameOrg, Ana_PMMP}. These evaluations are valuable for establishing whether particular migration mechanisms operate as intended, but they provide limited evidence about whether developers and organisations can successfully execute the migration process. Comparatively less evidence addresses migration effort, developer understanding, decision-making, usability of migration support, organisational coordination, maintainability, or the effectiveness of migration processes under realistic constraints. This distinction is important because migration involves not only technical transformation but also organisational decision-making and coordination across systems and stakeholders \cite{naether_migrating_2024}. Consequently, current evidence provides stronger support for the technical feasibility of individual migration activities than for the operational effectiveness of end-to-end migration.

The framework-tool divide also highlights a broader coordination gap between organisational migration planning and developer-level execution. Frameworks can establish migration priorities, governance structures, and decision processes, while tools can identify vulnerable components, assist implementation, or refactor code \cite{Ana_MigFrDepen, Ana_MigFrameOrg, Ana_PMMP, Ana_rattanavipanon_toolchain_2025, Ana_android_Migr, Ana_ccPASTpqc, Ana_legacyToQuantumConvert}. However, limited evidence demonstrates how outputs from one layer are systematically translated into activities at another. For example, a cryptographic inventory may identify high-priority assets without directly generating actionable developer tasks; an algorithm-selection decision may identify a preferred mechanism without defining corresponding implementation, testing, and deployment requirements; and a refactoring tool may modify code without providing evidence that the resulting changes satisfy the organisation's migration objectives. Effective migration therefore requires traceability between organisational requirements, cryptographic assets, migration decisions, implementation changes, verification results, and operational status. The absence of such traceability limits the ability to coordinate migration across organisational and technical boundaries.

\begin{tcolorbox}[
title=\textbf{Takeaway},
colback=gray!5,
colframe=black,
boxrule=0.5pt,
arc=0.5mm,
left=0.5mm,
right=0.5mm,
top=0.5mm,
bottom=0.5mm
]
Existing PQC migration support provides individual migration tasks rather than an integrated migration pathway. The key gap is the lack of empirically validated approaches that connect what needs to be migrated across the SDLC with how developers and organisations can coordinate planning, implementation, verification, deployment, and maintenance in practice.
\end{tcolorbox}

\subsection{Developer Education}
\label{DisEduReady}

The reviewed developer education approaches demonstrate that the literature has established a foundation for building PQC awareness and conceptual knowledge, but provides substantially less support for developing the practical competencies required for secure PQC integration. Existing approaches introduce learners to PQC algorithms, standardisation, protocols, quantum threats, and migration concepts \cite{Ana_borrelli_designing_2024, Ana_borrelli_towards_2025, Ana_edComScSociety, Ana_PQCIP, Ana_EducModule}. However, the coverage of these topics does not necessarily translate into the ability to perform PQC-related software engineering tasks. In particular, the reviewed approaches provide limited evidence of practical training skills such as secure algorithm and API selection, implementation, and testing. This distinction is important because PQC integration requires developers not only to understand PQC algorithms, but also to apply them correctly within existing software systems and development processes \cite{anonymous_pqc_sok_2026}. Thus, the current education landscape appears to address the knowledge acquisition dimension of PQC integration more strongly than the development of implementation competence.

The uniform alignment of the reviewed approaches with NIST CSF v2 Protect-Awareness and Training (PR.AT) further indicates a narrow cybersecurity orientation in existing PQC education. Although some approaches introduce quantum threat analysis, vulnerability assessment, and migration concepts \cite{Ana_edComScSociety, Ana_PQCIP}, these activities are primarily presented as educational content rather than as embedded competencies. None of the reviewed approaches explicitly extends their educational objectives to Detect, Respond, or Recover activities, while the Identify function is not explicitly represented despite the relevance of asset discovery and quantum-vulnerability assessment to PQC migration. Consequently, existing education approaches tend to frame PQC primarily as a problem of awareness and knowledge acquisition rather than as a lifecycle-oriented software security capability. 

The emergence of adaptive and game-based approaches indicates an increasing interest in moving beyond conventional content delivery toward more interactive learning experiences \cite{Ana_PQCLabEdu, Ana_abdelhamid_wip_2024}. These approaches are potentially valuable because PQC engineering involves decision-making and practical application rather than passive knowledge acquisition. However, the available evidence is not yet sufficient to conclude that interactive approaches produce superior PQC learning outcomes or, more importantly, improved secure implementation capability. The existing evaluations primarily establish prototype functionality, learner experience, or short-term educational outcomes \cite{Ana_PQCLabEdu, Ana_abdelhamid_wip_2024}. Therefore, the literature currently demonstrates the feasibility of interactive PQC education more strongly than its effectiveness in developing sustained software engineering competence.

Furthermore, existing evaluations mainly involve academic learners rather than practising software engineers \cite{Ana_borrelli_towards_2025, Ana_borrelli_designing_2024, Ana_edComScSociety, Ana_PQCLabEdu, Ana_abdelhamid_wip_2024}. They primarily assess immediate learning experiences or short-term outcomes, with limited evidence of knowledge transfer to realistic PQC engineering tasks, long-term retention, or changes in secure development practices. Therefore, evaluating an educational intervention does not necessarily demonstrate its effectiveness for PQC integration. Stronger evaluation should assess practical capabilities, including identifying vulnerable cryptographic dependencies, selecting and securely using PQC mechanisms, detecting implementation weaknesses, integrating PQC into existing systems, and performing migration tasks, with evidence of sustained capability over time.

The predominance of academic learners also limits the extent to which current findings can be generalised to professional software engineering environments. PQC integration in practice occurs under constraints that are not necessarily represented in educational settings, such as interoperability requirements, performance constraints, and organisational policies. The single approach involving IT professionals provides some evidence beyond academic populations \cite{Ana_PQCIP}, but this remains insufficient to establish how PQC education affects developers operating within realistic software development contexts. Consequently, there is a need for evaluation designs that involve practising developers and assess the quality of representative PQC implementation and migration tasks rather than relying primarily on perceptions or immediate learning outcomes.

\begin{tcolorbox}[
title=\textbf{Takeaway},
colback=gray!5,
colframe=black,
boxrule=0.5pt,
arc=0.5mm,
left=0.5mm,
right=0.5mm,
top=0.5mm,
bottom=0.5mm
]
Existing PQC education approaches provide foundational knowledge and awareness rather than integrated developer competence. The key gap is the lack of empirically validated approaches that connect PQC knowledge with hands-on decision-making, secure implementation, testing, migration, deployment, and maintenance while evaluating whether these capabilities transfer to realistic software engineering practice.
\end{tcolorbox}

\subsection{Cross-Cutting Gaps}
\label{DisOverallReady}

Collectively, the findings reveal that PQC implementation, migration, and developer education have evolved as largely separate streams with limited integration across the integration process. Implementation approaches focus on technical mechanisms and validation, migration approaches on planning and transformation, and education approaches on awareness and knowledge. Limited evidence shows how these approaches can be coordinated to translate organisational integration goals into developer actions and support subsequent activities. In particular, migration approaches can identify cryptographic assets, priorities, and replacement strategies, while implementation approaches support coding and testing, but the linkage between these outputs and developer-level execution remains unclear. The common gap is therefore fragmentation across organisational, technical, and human activities, rather than the absence of individual approaches. This fragmentation limits traceability across the PQC integration process, from asset identification and decision-making through implementation, testing, deployment, and maintenance \cite{Ana_MigFrDepen, Ana_MigFrameOrg, Ana_PMMP, Ana_decisionAlgo, naether_migrating_2024}.

Next, there is a gap between technical feasibility and operational deployment. Both implementation and migration studies demonstrate that individual mechanisms can be implemented or evaluated under controlled conditions, but provide limited evidence of sustained use in realistic software environments. In reality, legacy dependencies, interoperability, and organisational responsibilities can introduce constraints not captured by laboratory evaluations. Thus, current evidence establishes technical feasibility more strongly than reliable operational use \cite{nist_fips_203_2024, nist_fips_204_2024, nist_fips_205_2024}.

A further cross-cutting gap concerns the evaluation and assurance of PQC-integrated software. Existing evaluations are often ad hoc and focus on performance, functional correctness, protocol conformance, and selected security properties, with limited systematic software verification and validation. NIST IR 8397 provides general software verification and testing guidance \cite{nist_ir_8397_2021}, while NIST PQC evaluation criteria focus mainly on cryptographic security, algorithmic properties, and cost \cite{nist_pqc_cost_evaluation_criteria, nist_pqc_security_evaluation_criteria}. Limited guidance exists for assessing security after PQC integration into software systems, where risks may arise from non-constant-time execution, side-channel leakage, insecure parameter handling or randomisation, API misuse, incorrect configuration, and integration vulnerabilities \cite{toruan2026security, Ana_hekkala_implementing_2023}. Thus, a gap remains between evaluating PQC security in isolation and establishing that its software integration is securely implemented, verified, validated, and maintained.

Education approaches reveal a complementary developer competency gap. Existing interventions provide useful PQC awareness and conceptual knowledge, but provide less evidence of competence in realistic implementation and migration tasks that support PQC integration. The predominance of academic learners and short-term evaluations further limits evidence of knowledge transfer, retention, and changes in secure development practice \cite{Ana_borrelli_towards_2025, Ana_borrelli_designing_2024, Ana_PQCLabEdu, Ana_abdelhamid_wip_2024}. Consequently, technical and organisational support may remain difficult to operationalise without corresponding developer competencies.

Taken together, these findings indicate that the principal gap is not a lack of PQC algorithms, implementation approaches, migration approaches or educational interventions, but the absence of an empirically validated socio-technical pathway connecting organisational decisions, developer competencies, and secure PQC integration activities. The literature demonstrates individual capabilities more strongly than their coordinated and sustained use across the software development lifecycle.

\subsection{Recommendations}
\label{DisResearchD}

Based on the findings presented and discussed in Sections~\ref{Resul} and~\ref{Discuss}, four priorities emerge for advancing PQC integration research and practice.

\textbf{1) Integrated lifecycle support:} The fragmentation of implementation and migration support across the SDLC indicates a need for stronger coordination throughout PQC integration. Future approaches should connect activities from cryptographic asset identification and decision-making through implementation, testing, deployment, and maintenance, with traceability between organisational decisions and developer activities.

\textbf{2) PQC-specific software assurance:} The limited coverage and ad hoc nature of PQC integration testing indicates that existing evaluation guidance is insufficient for PQC-integrated software. Standards bodies and researchers should develop PQC-specific guidance addressing concerns such as constant-time execution, side-channel leakage, parameter handling, API misuse, configuration, interoperability, and systematic software QA, testing, verification, and validation.

\textbf{3) Competency-oriented developer education:} The predominance of conceptual learning indicates that existing education provides insufficient preparation for practical PQC integration. Training should therefore emphasise applied competencies through realistic implementation and migration tasks, including algorithm and API selection, secure implementation, testing, verification, and migration decision-making, with evaluation based on demonstrated task performance.

\textbf{4) Realistic evaluation and deployment evidence:} The predominance of controlled experimental evidence indicates limited understanding of PQC integration in operational settings. Future studies should, where feasible, involve practising developers and realistic codebases, dependencies, deployment constraints, and longer-term maintenance to establish whether proposed approaches support secure and sustainable PQC integration in practice.

\section{Conclusion}
\label{Conclu}

This SoK examined the landscape of PQC integration in software development, focusing on implementation, migration, and developer education. The findings show that the literature has established important implementation, migration, and educational approaches, but these remain fragmented across the PQC integration process. Implementation approaches primarily demonstrate technical mechanisms, migration approaches support selected planning and transformation activities, and education approaches provide foundational PQC knowledge. However, limited evidence demonstrates how these approaches can be coordinated across the SDLC and translated into sustained practice.

The analysis further identified four cross-cutting gaps: limited traceability between organisational migration decisions and developer activities, stronger evidence of technical feasibility than operational capability, insufficient guidance for testing and assuring PQC-integrated software, and limited evidence that existing education develops practical developer competence. These gaps indicate that secure PQC integration requires more than cryptographic algorithms or individual implementation and migration tools. It requires an integrated socio-technical approach that connects organisational decision-making, developer capabilities, secure engineering practices, and continuous assurance.

Future research should therefore focus on lifecycle-integrated PQC support, PQC-specific software testing and assurance guidance, competency-oriented developer education, and empirical evaluation in realistic software engineering environments. Addressing these gaps can move PQC research beyond demonstrating isolated support toward establishing evidence-based approaches for secure and sustainable PQC integration in practice.

\section*{AI Usage}
AI-supported tools were used only as writing assistants to improve grammar, clarity, and readability. The authors independently verified all generated text and remain fully responsible for the methodology, literature analysis, findings, and conclusions. No AI tools were used for literature selection, evidence extraction, analysis, or research decision-making.

\section*{Ethical Considerations}

\begin{itemize}
    \item \textbf{Stakeholders:} The work is relevant to software developers, researchers, organisations, standards bodies, API and library maintainers, educators, and end users. The research team is also a stakeholder because methodological choices affect the interpretation of the evidence.
    
    \item \textbf{Ethical Principles:} The analysis follows the USENIX Security stakeholder framework and considers beneficence, respect for persons, justice, and respect for law and public interest.
    
    \item \textbf{Potential Benefits:} This SoK provides an evidence-based overview of PQC integration, highlighting current strengths, limitations, and research gaps. The findings can support researchers, practitioners, educators, and standards bodies in improving secure PQC integration mechanisms.
    
    \item \textbf{Potential Harms:} Readers may over-interpret the findings as evidence that existing PQC technologies are unsuitable or, conversely, deploy them without sufficient security practices. Publication bias may also limit the completeness of the findings. No human participants or sensitive data were involved, and no offensive security activities were conducted.
    
    \item \textbf{Harm Mitigation:} We use a transparent review methodology and established assessment criteria. We distinguish research preparedness from the security of PQC algorithms and avoid claims beyond the available evidence. Limitations, including incomplete SDLC coverage and limited empirical validation, are explicitly acknowledged.
    
    \item \textbf{Ethical Decision:} The anticipated benefits outweigh the potential risks. The study does not disclose vulnerabilities or enable attacks; instead, it supports evidence-based PQC integration by identifying areas requiring further research, guidance, education, and validation.
\end{itemize}

\section*{Open science}

This paper presents a SoK on the landscape of PQC integration in software development. The study is based entirely on a systematic review and evidence synthesis of publicly available literature and does not involve the development of software, collection of experimental data, user studies, or empirical system evaluations.

\begin{itemize}
    \item \textbf{Artifact Categories and Availability: } This work does not produce research artifacts such as source code, datasets, executable software, scripts, or experimental benchmarks. The contribution consists of a systematic synthesis and evidence-based assessment of published research on PQC integration approaches. The literature analysed in this study comprises peer-reviewed publications and publicly available standards and guidance documents, all of which are cited in the bibliography and accessible through standard academic repositories or official publication channels. The taxonomy of PQC integration approaches, thematic classifications, and evidence mapping methodology are described within the paper. Consequently, no additional computational artifacts are required to understand, evaluate, or reproduce the analytical contribution.
    \item \textbf{Reproducibility: } The findings presented in this SoK can be independently reproduced by following the systematic review methodology described in the paper. This includes conducting literature searches using the reported search strategy and eligibility criteria, extracting evidence from the included studies, and evaluating the identified approaches using the mapping dimensions derived from authoritative guidance. All information necessary to understand, verify, and evaluate the contributions of this work is contained within the paper and its publicly accessible references.
\end{itemize}

\bibliographystyle{plain}
\bibliography{SoK_NIST_Main}

\appendix
\section{Supplementary Information on the Research Methodology}
\label{appendix:Methodology}

\subsection{PICOC Table}
\label{picocTable}

\begin{table*}[!t]
\centering
\caption{Application of the PICOC framework to develop the search strategy for RQ1.}
\label{tab:PICOC}
\small
\begin{tabular}{p{1.5cm}p{3.5cm}p{9cm} }
\toprule
\textbf{Concept} & \textbf{Application} & \textbf{Representative Keywords} \\
\midrule
Population &
Software development &
software development, software engineering, software system \\

Intervention &
PQC integration approaches &
integration, migration, implementation, framework, guideline, library, tool, education, awareness \\

Comparison &
\multicolumn{2}{c}{Not applicable}\\

Outcome &
PQC integration approaches &
solution*, recommendation*, approach*, method*, framework*, technique*, strateg*, mitigation, resolution, remed*, intervention*, best practice* \\

Context &
PQC integration in software development &
PQC, Post-Quantum Cryptography, quantum-resistant cryptography, quantum-safe cryptography \\
\bottomrule
\end{tabular}
\end{table*}
\FloatBarrier

The PICOC framework was applied to systematically define the scope of RQ1 \textit{(What approaches have been proposed or implemented to support the integration of PQC in software systems?)} and construct the search strategy. The Population identifies the target domain of investigation, focusing on software systems and development where PQC integration is required. The Intervention captures the different approaches proposed or implemented to support PQC integration, including frameworks, guidelines, tools, libraries, migration strategies, and educational initiatives. Since RQ1 aims to identify and classify existing PQC integration approaches rather than compare specific interventions, the Comparison element is not applicable. The Outcome represents the expected contributions of these approaches, such as solutions, methods, recommendations, techniques, and best practices that facilitate PQC integration. The Context defines the broader research setting, focusing on the integration of PQC in software systems. Based on these PICOC elements, relevant keywords and their synonyms were identified and combined using Boolean operators to formulate the search strategy.

\subsection{Inclusion and Exclusion Criteria}
\label{InEXList}

Clear study selection criteria are essential to ensure the relevance and quality of an SLR \cite{booth_systematic_2016}. Accordingly, we formulated explicit inclusion (Inc) and exclusion (Exc) criteria to guide the selection process, as outlined below:

\begin{itemize}
  \item \textbf{Inclusion Criteria}
      \begin{itemize}
       \item Inc1: Research addressing PQC integration in software development
        \item Inc2: Publications proposing, evaluating, or discussing methodologies, frameworks, guidelines, tools, libraries, or educational interventions for PQC integration.
        \item Inc3: Peer-reviewed journal articles, conference papers, or workshop papers.
        \item Inc4: Publications written in English.
        \item Inc5: Publications from January 2020 to June 2026.\footnote{Two studies published in 2017 and 2019 were included as controlled exceptions through backward and forward snowballing due to their relevance. All other included studies satisfied the predefined publication period.}
      \end{itemize}

  \item \textbf{Exclusion Criteria}
      \begin{itemize}
        \item Exc1: Publications unrelated to PQC integration in software development.
        \item Exc2: Non-peer-reviewed literature (e.g., blog posts, opinion articles, white papers, and technical reports).
        \item Exc3: Duplicate or substantially overlapping publications.
        \item Exc4: Publications without accessible full text.
        \item Exc5: Non-English publications.
      \end{itemize}
\end{itemize}

%%%%%%%%%%%%%%%%%%%%%%%%%%%%%%%%%%%%%%%%%%%%%%%%%%%%%%%%%%%%%%%%%%%%%%%%%%%%%%%%
\end{document}